\documentclass[conference]{IEEEtran}
\IEEEoverridecommandlockouts
\usepackage{cite}
\usepackage{amsmath,amssymb,amsfonts}
\usepackage{algorithmic}
\usepackage{graphicx}
\usepackage{textcomp}
\usepackage{xcolor}
\usepackage{cite}
\usepackage{subfigure}

\author{
    \IEEEauthorblockN{M. Koumans and J.L.M. van Mechelen}
    \IEEEauthorblockA{
    Electrical Engineering, Eindhoven University of Technology, Eindhoven, Netherlands}
}

\def\BibTeX{{\rm B\kern-.05em{\sc i\kern-.025em b}\kern-.08em
    T\kern-.1667em\lower.7ex\hbox{E}\kern-.125emX}}

\begin{document}

\title{Physics-based AI methodology for\\ Material Parameter Extraction from Optical Data}

\maketitle
\begin{abstract}
We report on a novel methodology for extracting material parameters from spectroscopic optical data using a physics-based neural network. The proposed model integrates classical optimization frameworks with a multi-scale object detection framework, specifically exploring the effect of incorporating physics into the neural network. We validate and analyze its performance on simulated transmission spectra at terahertz and infrared frequencies. Compared to traditional model-based approaches, our method is designed to be autonomous, robust, and time-efficient, making it particularly relevant for industrial and societal  applications.
\end{abstract}

\section{Introduction}

State-of-the-art signal processing related to material parameter extraction primarily relies on model-based and data-driven approaches. In model-based methods, the inverse problem of determining material properties from an observable, such as a transmission spectrum, is addressed through least-squares optimization of a descriptive model \cite{van_mechelen_stratified_2014}. Typically, this model is based on the Fresnel equations implemented through the transfer matrix method, having a predefined set of fitting parameters. In contrast, data-driven approaches leverage conventional machine learning techniques to optimize a self-selected model to best describe the given dataset. The underlying principles of these data-driven methods range from identifying statistical correlations related to the output to detecting patterns—and thereby extracting material properties.


The model-based approach has proven effective in well-controlled environments for planar layered systems with minimal variability~\cite{thickness_sensing}. However, as variability increases, the number of fitting parameters becomes a fundamental challenge. For instance, in industrial applications such as paint layer inspection, more than four layers each have at least five fitting parameters. Allowing all parameters to vary without human intervention makes the problem underdetermined and prone to erroneous outputs. Additionally, real-world objects often have complex, non-planar shapes and exist in variable environments, further complicating the modelling process. Rather than addressing these challenges, advancements in model-driven methods aim at refining existing convergence optimisation algorithms~\cite{spectral_projected_gradient}.

Data-driven approaches, on the other hand, are inherently flexible but require extensive training. As a result, they are considered a promising solution for extracting material parameters from real-world samples and for applications with high material variability. However, these methodologies are mostly applied as black boxes and are susceptible to overfitting. To gain deeper insight, we recently attempted to validate the activation functions of a convolutional neural network, applied to a time-domain signal, by comparing them to the Fresnel equations. Despite the apparent correspondence, even a slight modification of the dataset beyond the training set led to poor performance \cite{sr_koumans}.


Recently, hybrid approaches, where prior knowledge is incorporated into data-driven signal processing, have gained traction, improving effectiveness by mitigating model overfitting and enhancing predictive capabilities \cite{PIML,  chen_hybrid_2021}. Given the well-established equations governing light-matter interactions in multilayer systems, we propose a hybrid approach that embeds this physics within a data-driven framework.

\begin{figure}
    \centering
    \includegraphics[width=.9\linewidth]{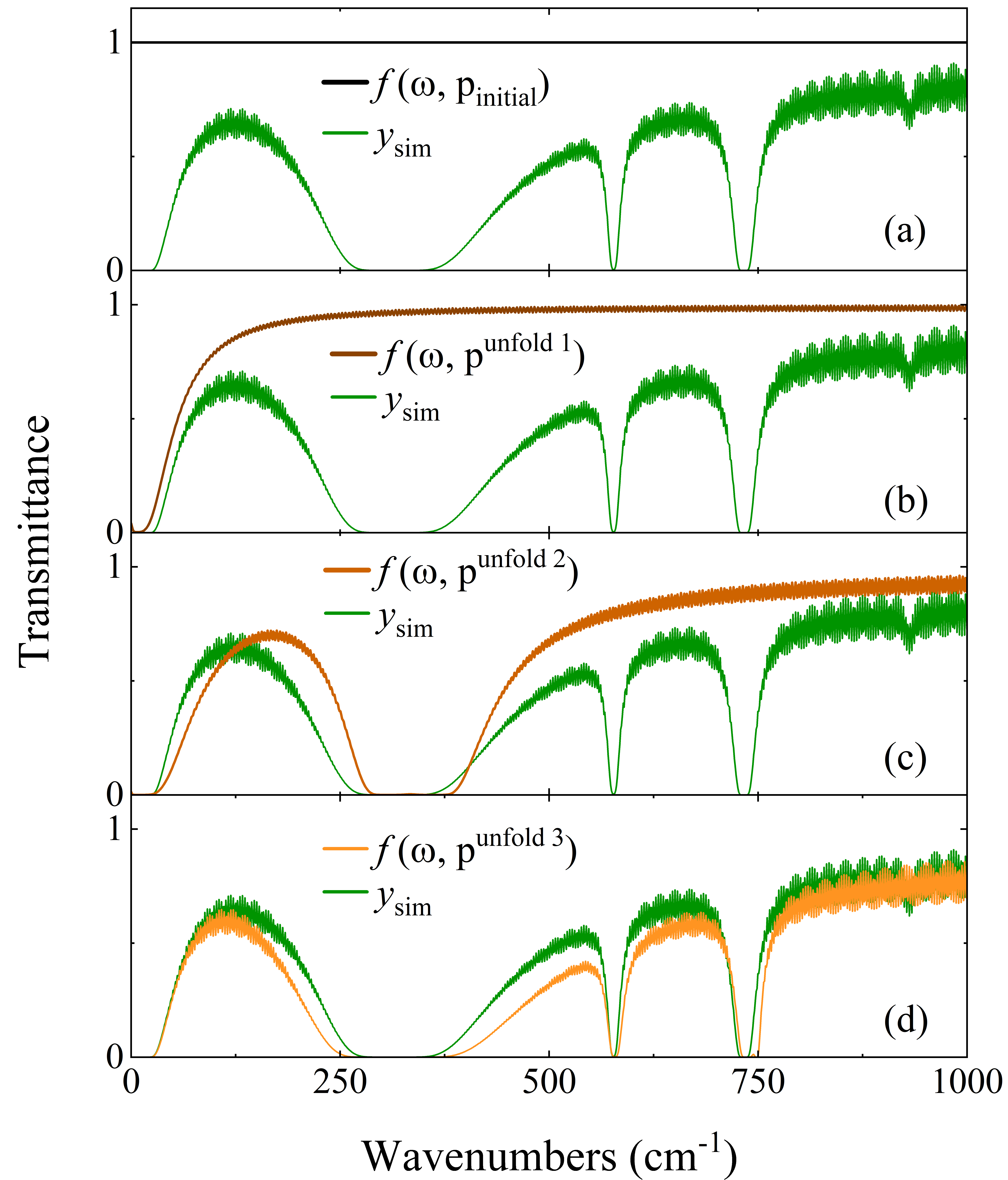}
    \caption{Ground truth signal $y_{\text{sim}}$ and fitted signal $f(\omega, p^{[\ell]})$ using our physics-based AI approach, for $p$ containing (a) the initial values, and the updated values after (b) one unfold, (c) two unfolds, and (d) three unfolds.}    
    \label{fig:1}
\end{figure}

\section{Results}
We propose a physics-based deep learning model that autonomously extracts optical material properties from an optical spectrum. The model fuses classical optimization with a multi-scale object detection framework. The traditional iterative optimization scheme used in nonlinear fitting is unfolded as a fixed-sized forward model, where each update step corresponds to an unfold layer $\ell$ of the network. At each unfold $\ell$, the intermediate fitted signal $f(\omega, p^{[\ell]})$ is determined by the system parameters $p^{[\ell]}$ through the system model $F_{\text{system}}(\cdot)$, expressed as a transfer matrix method with a Drude-Lorentz parameterization of the dielectric function $\epsilon(\omega)$. The ground truth input spectrum $y_{\text{sim}}$, together with $f(\omega,p^{[\ell]})$, is processed based on the principles of single-shot detectors. The single-shot detector outputs candidate oscillator parameters, which are filtered based on their objectness score. If the score meets a predefined threshold, the parameters are added to $p^{[\ell]}$. As a result, output $\hat{p}$ consists of predefined system parameters: $\epsilon_{\infty}$, the Drude components $\Omega_{p}$ and $\Gamma$, along with a variable number of parameters $\omega_{0, i}$, $\omega_{p, i}$, and $\gamma_{i}$ for each Lorentz oscillator $i$.

We apply our hybrid model to a simulated transmission spectrum $y_{\text{sim}}$ at THz and infrared frequencies $\omega$, ranging from 0 to 1000 cm$^{-1}$ (Fig. \ref{fig:1}a). $y_{\text{sim}}$ contains a Drude component, four absorptions with Lorentzian dispersion, and Fabry-Pérot oscillations due the condition $\omega\lesssim \sqrt{\epsilon(\omega)}d$, where $d$ is the sample thickness. Each spectrum is modeled as a three-layer system, where both the substrate and superstrate are air with $\epsilon_{\text{air}}(\omega) = 1 + 0j$, and the sample layer is characterized by $\epsilon(\omega)$ and $d$. The hybrid model has a total number of 30,000 trainable parameters and consists of three unfolds, each using the same feature detection block (backbone). Figure \ref{fig:1} shows the model’s output at each unfold. Given the starting values of $p^{[\ell]}$, initially $f(\omega, p^{[\ell]}) = 1$ for all $\omega$. The model parameters are updated through $\Delta p^{[\ell]}$ at each step, based on the learned relation between $y_{\text{sim}}$ and $f(\omega, p^ {[\ell]})$. In the first unfold, $f(\omega, p^{[1]})$ is primarily sensitive to the Drude contribution and, to a lesser extent, the Fabry-Pérot oscillations. In the second unfold, $f(\omega, p^{[2]})$ refines both aspects while also adapting to the strongest Lorentz oscillator. In the final unfold, $f(\omega, p^{[3]})$ fits to the weaker Lorentz oscillators and is fine-tuned to the Fabry-Pérot oscillations. Notably, this update procedure differs from a manual model-based approach, where all oscillators are first positioned and tuned before addressing the Fabry-Pérot oscillations.

\begin{figure}
    \centering
    \includegraphics[width=\linewidth]{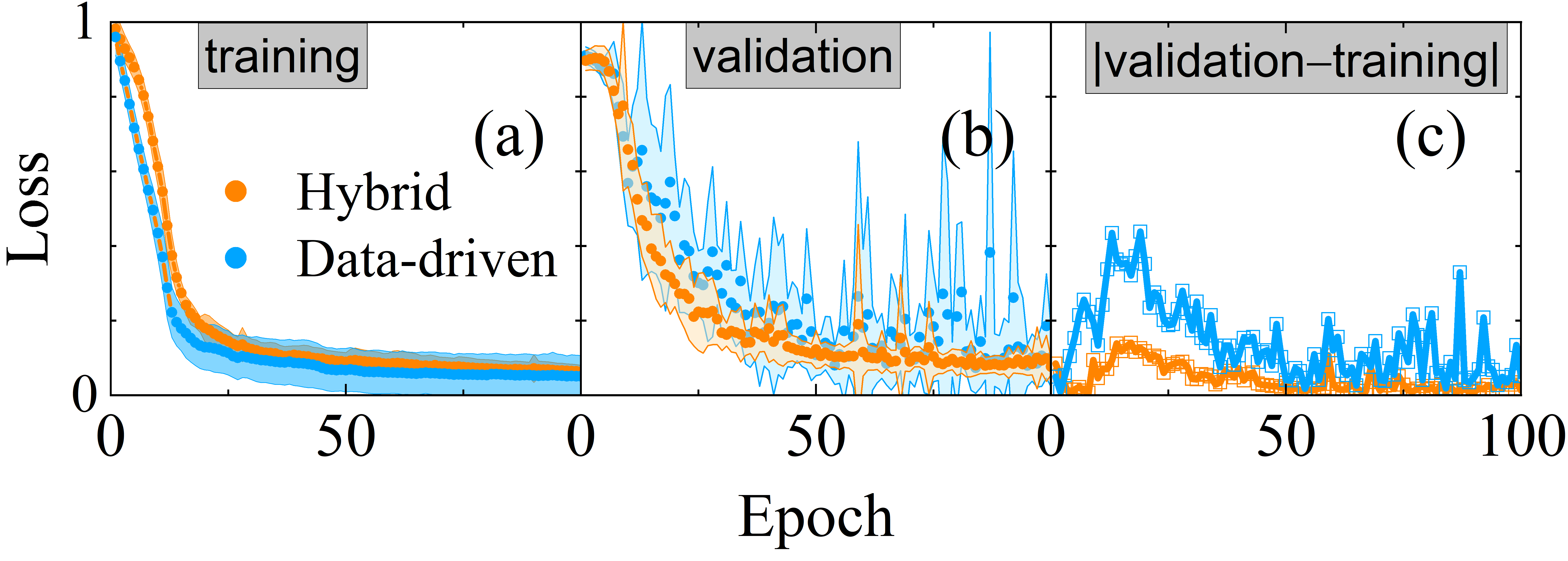}
    \caption{Loss related to (a) training and (b) validation of a hybrid approach (orange) and purely data-driven approach (blue). (c) Performance variability evaluated by $|\mathcal{L}_{\text{val}} - \mathcal{L}_{\text{train}}|$ for each epoch. }    
    \label{fig:2}
\end{figure}

We validate and analyze the performance of the proposed model using a large set of simulated transmission spectra. In addition, we also train a purely data-driven approach, unlinked to the transfer matrix method. Both models are trained for 100 epochs across 10 independent runs. In each run, we generate 8192 spectra for training and 1024 spectra each for validation and testing. The trainable parameters are randomly initialized according to a uniform distribution. By training over multiple runs, we assess the stability of the training process. The models are optimized using a custom loss function that accounts for optical parameter values, system parameters, and objectness scores. The training loss $\mathcal{L}_{\text{train}}$ shows that the purely data-driven approach exhibits slightly faster and better convergence on the training set compared to the hybrid model (Fig.~\ref{fig:2}a). However, its convergence stability is significantly lower. In contrast, the validation loss $\mathcal{L}_{\text{val}}$ fluctuates considerably for the unconstrained data-driven model, while the hybrid model exhibits more consistent behaviour (Fig.~\ref{fig:2}b). The generalization gap at each epoch, defined as $\Delta_{\text{gen}} = |\mathcal{L}_{\text{val}} - \mathcal{L}_{\text{train}}|$ serves as a quantitative measure of model overfitting. $\Delta_{\text{gen}}$ exhibits a characteristic bell shape in the initial phase for both methodologies (Fig.~\ref{fig:2}c). However, the uninformed model shows a larger and less stable $\Delta_{\text{gen}}$, indicating greater susceptibility to overfitting. This suggests that the inductive bias introduced in the hybrid model constrains its convergence, leading to improved generalization.

\section{Conclusion}
We have proposed a hybrid physics-based AI approach for extracting material parameters from optical data. The model enables autonomous parameter retrieval from transmission spectra within the trained parameter domain. Furthermore, our results demonstrate that incorporating physics directly into our deep learning framework leads to a more challenging training process but results in a more generalizable model. Notably, the proposed model is flexible with respect to the selected spectral range, the number of oscillators, and the parameterization of the dispersion. These findings pave the way for a new paradigm in autonomous signal processing for optical spectroscopic data.

\bibliographystyle{IEEEtran}
\bibliography{IRMMW-THz}

\begin{thebibliography}{1}
\providecommand{\url}[1]{#1}
\csname url@samestyle\endcsname
\providecommand{\newblock}{\relax}
\providecommand{\bibinfo}[2]{#2}
\providecommand{\BIBentrySTDinterwordspacing}{\spaceskip=0pt\relax}
\providecommand{\BIBentryALTinterwordstretchfactor}{4}
\providecommand{\BIBentryALTinterwordspacing}{\spaceskip=\fontdimen2\font plus
\BIBentryALTinterwordstretchfactor\fontdimen3\font minus \fontdimen4\font\relax}
\providecommand{\BIBforeignlanguage}[2]{{%
\expandafter\ifx\csname l@#1\endcsname\relax
\typeout{** WARNING: IEEEtran.bst: No hyphenation pattern has been}%
\typeout{** loaded for the language `#1'. Using the pattern for}%
\typeout{** the default language instead.}%
\else
\language=\csname l@#1\endcsname
\fi
#2}}
\providecommand{\BIBdecl}{\relax}
\BIBdecl

\bibitem{van_mechelen_stratified_2014}
J.~L.~M. van Mechelen, A.~B. Kuzmenko, and H.~Merbold, ``Stratified dispersive model for material characterization using terahertz time-domain spectroscopy,'' \emph{Optics Letters}, vol.~39, no.~13, pp. 3853--3856, 2014.

\bibitem{thickness_sensing}
\BIBentryALTinterwordspacing
J.~L.~M. Van~Mechelen, A.~Frank, and D.~J. H.~C. Maas, ``\BIBforeignlanguage{en}{Thickness sensor for drying paints using {THz} spectroscopy},'' \emph{\BIBforeignlanguage{en}{Optics Express}}, vol.~29, no.~5, p. 7514, Mar. 2021. [Online]. Available: \url{https://opg.optica.org/abstract.cfm?URI=oe-29-5-7514}
\BIBentrySTDinterwordspacing

\bibitem{spectral_projected_gradient}
\BIBentryALTinterwordspacing
E.~G. Birgin, J.~M. Martínez, and M.~Raydan, ``\BIBforeignlanguage{en}{Spectral {Projected} {Gradient} {Methods}: {Review} and {Perspectives}},'' \emph{\BIBforeignlanguage{en}{Journal of Statistical Software}}, vol.~60, no.~3, 2014. [Online]. Available: \url{http://www.jstatsoft.org/v60/i03/}
\BIBentrySTDinterwordspacing

\bibitem{sr_koumans}
\BIBentryALTinterwordspacing
M.~Koumans, D.~Meulendijks, H.~Middeljans, D.~Peeters, J.~C. Douma, and D.~Van~Mechelen, ``\BIBforeignlanguage{en}{Physics-assisted machine learning for {THz} time-domain spectroscopy: sensing leaf wetness},'' \emph{\BIBforeignlanguage{en}{Scientific Reports}}, vol.~14, no.~1, p. 7034, Mar. 2024. [Online]. Available: \url{https://www.nature.com/articles/s41598-024-57161-4}
\BIBentrySTDinterwordspacing

\bibitem{PIML}
\BIBentryALTinterwordspacing
M.~Raissi, P.~Perdikaris, and G.~Karniadakis, ``\BIBforeignlanguage{en}{Physics-informed neural networks: {A} deep learning framework for solving forward and inverse problems involving nonlinear partial differential equations},'' \emph{\BIBforeignlanguage{en}{Journal of Computational Physics}}, vol. 378, pp. 686--707, Feb. 2019. [Online]. Available: \url{https://linkinghub.elsevier.com/retrieve/pii/S0021999118307125}
\BIBentrySTDinterwordspacing

\bibitem{chen_hybrid_2021}
\BIBentryALTinterwordspacing
X.~Chen, Z.~Yao, S.~Xu, A.~S. McLeod, S.~N. Gilbert~Corder, Y.~Zhao, M.~Tsuneto, H.~A. Bechtel, M.~C. Martin, G.~L. Carr, M.~M. Fogler, S.~G. Stanciu, D.~N. Basov, and M.~Liu, ``\BIBforeignlanguage{en}{Hybrid {Machine} {Learning} for {Scanning} {Near}-{Field} {Optical} {Spectroscopy}},'' \emph{\BIBforeignlanguage{en}{ACS Photonics}}, vol.~8, no.~10, pp. 2987--2996, Oct. 2021. [Online]. Available: \url{https://pubs.acs.org/doi/10.1021/acsphotonics.1c00915}
\BIBentrySTDinterwordspacing

\end{thebibliography}

\end{document}